\documentclass[sigconf]{acmart}

\acmConference[ICSE 2024]{46th International Conference on Software Engineering}{April 2024}{Lisbon, Portugal}
\AtBeginDocument{%
  \providecommand\BibTeX{{%
    \normalfont B\kern-0.5em{\scshape i\kern-0.25em b}\kern-0.8em\TeX}}}

\acmPrice{15.00}
\acmISBN{978-1-4503-XXXX-X/18/06}

\usepackage{nicematrix}
\usepackage{tikz}
\usepackage{amsmath}
\usepackage{caption}
\usepackage{multirow}
\usepackage{comment}
\usepackage{float}
\usepackage{array}
\usepackage{tabularx}
\usepackage{blindtext}
\renewenvironment{quote}
  {\list{}{\rightmargin=0.27in \leftmargin=0.27in}%
   \item\relax}
  {\endlist}
\usepackage{svg}
\usepackage{dirtytalk}

\usepackage{dcolumn}
\usepackage{hyperref}
\usepackage{breakurl}

\graphicspath{ {./img/} }
       
\newenvironment{myparticipantquote}[2]
{\say{\textit{#1}} (#2)} 

\copyrightyear{2024} 
\acmYear{2024} 
\setcopyright{rightsretained} 
\acmConference[CHASE '24]{2024 IEEE/ACM 17th International Conference on Cooperative and Human Aspects of Software Engineering }{April 14--15, 2024}{Lisbon, Portugal}
\acmBooktitle{2024 IEEE/ACM 17th International Conference on Cooperative and Human Aspects of Software Engineering (CHASE '24), April 14--15, 2024, Lisbon, Portugal}\acmDOI{10.1145/3641822.3641872}
\acmISBN{979-8-4007-0533-5/24/04}

\begin{document}

\title{Understanding the Career Mobility of Blind and Low Vision Software Professionals}

\author{Yoonha Cha}
\email{yoonha.cha@uci.edu}
\affiliation{%
  \institution{University of California, Irvine}
  \streetaddress{Donald Bren Hall}
  \city{Irvine}
  \state{CA}
  \country{USA}
  \postcode{92617}
  }

\author{Victoria Jackson}
\email{vfjackso@uci.edu}
\affiliation{%
  \institution{University of California, Irvine}
  \streetaddress{Donald Bren Hall}
  \city{Irvine}
  \state{CA}
  \country{USA}
  \postcode{92617}
  }

\author{Isabela Figueira}
\email{i.figueira@uci.edu}
\affiliation{%
  \institution{University of California, Irvine}
  \streetaddress{Donald Bren Hall}
  \city{Irvine}
  \state{CA}
  \country{USA}
  \postcode{92617}
  }

\author{Stacy M. Branham}
\email{sbranham@uci.edu}
\affiliation{%
  \institution{University of California, Irvine}
  \streetaddress{Donald Bren Hall}
  \city{Irvine}
  \state{CA}
  \country{USA}
  \postcode{92617}
  }

\author{André van der Hoek}
\email{andre@ics.uci.edu}
\affiliation{%
  \institution{University of California, Irvine}
  \streetaddress{Donald Bren Hall}
  \city{Irvine}
  \state{CA}
  \country{USA}
  \postcode{92617}
  }

\renewcommand{\shortauthors}{Cha et al.}

\begin{abstract}
\textbf{Context}: Scholars in the software engineering (SE) research community have investigated career advancement in the software industry. Research topics have included how individual and external factors can impact career mobility of software professionals, and how gender affects career advancement. However, the community has yet to look at career mobility from the lens of accessibility. Specifically, there is a pressing need to illuminate the factors that hinder the career mobility of blind and low vision software professionals (BLVSPs).
\textbf{Objective}: This study aims to understand aspects of the workplace that impact career mobility for BLVSPs.
\textbf{Methods}: We interviewed 26 BLVSPs with different roles, years of experience, and industry sectors. Thematic analysis was used to identify common factors related to career mobility. 
\textbf{Results}: We found four factors that impacted the career mobility of BLVSPs: (1) technical challenges, (2) colleagues' perceptions of BLVSPs, (3) BLVSPs' own perceptions on managerial progression, and (4) BLVSPs' investment in accessibility at the workplace.
\textbf{Conclusion}: We suggest implications for tool designers, organizations, and researchers towards fostering more accessible workplaces to support the career mobility of BLVSPs.

\end{abstract}



\maketitle

\section{Introduction}

Today, in the United States, people with disabilities (PWD) comprise 27.8\% of the general population~\cite{cdc_disability_2023}, with Blind and Low Vision (BLV) adults making up 4.8\% of the general population~\cite{cdc_disability_2023}. Yet, only 1.7\% of the total software development workforce identified as being BLV, according to a survey conducted by Stack Overflow in 2022 on software industry workers~\cite{stackoverflow_2022}. While diversity is an important topic in SE research~\cite{menezes_diversity_2018, ACM_diversity, IEEE_diversity}, disability is not given equal attention in papers specifically focused on diversity, compared to other factors such as gender, age, and race/ethnicity~\cite{casey_include_2020, silveira_systematic_2019, rodriguez_perceived_2021, dagan_building_2023}. A systematic literature review conducted in 2019 found that only one out of 221 publications in SE and Agile Methodology venues regarding diversity focused on disability~\cite{silveira_systematic_2019}. 

Nonetheless, there is a small but growing body of research on BLV individuals in SE, with prior work documenting the challenges that BLV developers face, including the inaccessibility of Integrated Development Environments (IDEs)~\cite{albusays_eliciting_2016, albusays_interviews_2017, armaly_audiohighlight_2018, armaly_comparison_2018}, challenges when co-piloting during pair programming~\cite{filho_visual_2015, huff_workexp_2020, pandey_understanding_2021}, and comprehending and navigating project management tools such as Jira~\cite{filho_visual_2015, huff_workexp_2020, pandey_understanding_2021}. These insights characterize the inaccessibility of particular programming experiences for BLV developers. Yet, there is little indication in the literature of how these inaccessible experiences accumulate and affect career advancement for Blind and Low Vision Software Professionals (BLVSPs). 

Career advancement in the software industry has been investigated by SE researchers (e.g., ~\cite{tremblay_determinants_2002}). 
Recent research on careers in SE primarily focuses on how gender affects career mobility~\cite{ahuja_women_2002}. 
However, the role accessibility plays in career mobility of software professionals has been given little attention.
To address this gap, we conducted a qualitative study to answer the research question: \textit{What factors affect the career mobility of BLVSPs?} 

We conducted semi-structured interviews with 26 participants who identified as being blind or having low vision, worked or were working in a position within the software industry, and had at least one year of experience at the time of the interview. The study started out as an investigation of the accessibility of meetings as a factor in career mobility (e.g., \textit{Do you think the accessibility issues [in meetings] we discussed today have any impact on career advancement?}). The subsequent conversations that would ensue frequently concerned not just factors related to meetings impacting career mobility, but a much broader set of factors. This paper presents the results of our analysis of these factors and how they influence career mobility. 
We conducted a qualitative, inductive thematic analysis of our data~\cite{braun_using_2006} which revealed four factors that may influence BLVSPs' career mobility: technical challenges, colleagues' attitudes towards them, their own perceptions of managerial advancement, and their investment in workplace accessibility. Based on our findings, we discuss the implications for software development tools, organizations, and for researchers.

This study contributes the following: 
\begin{itemize}
    \item The documentation of accessibility of software tools used in software development meetings, and how the accessibility of the tools relates to the career mobility of BLVSPs.
    \item The documentation of factors that impact career mobility of BLVSPs.
    \item Implications for software development tools, organizations, and researchers.
    \item A call for further research on disability as part of diversity in SE.
\end{itemize}

\section{Background and Related Work}
In this section, we review relevant background of assistive technologies and the legal protections afforded to BLV workers in the United States (where the majority of our participants are based), the work experiences of BLV professionals, and career mobility in software engineering 

\paragraph{\textbf{Assistive Technologies}} 
Assistive technologies (AT) \cite{WHO_2023} are hardware devices or software applications designed to support PWD. Popular AT used by BLVSPs include screen readers and screen magnifiers. Screen readers (e.g., JAWS~\cite{jaws} and NVDA~\cite{NVDA}) enable BLVSPs to read the content on a screen either through a speech synthesizer or a braille display. Users control the screen reader to navigate, read, and interact with screen content. Screen magnifiers enable people with usable vision to zoom into a particular part of the screen to read the content. All of these ATs are used by BLVSPs to interact with text editors and IDEs while coding \cite{mealin_exploratory_2012}.

\paragraph{\textbf{Work Environment for People with Disabilities in the USA}}
 The Americans with Disabilities Act (ADA)~\cite{ADA.gov} guarantees PWD equal rights and protection from discrimination in many aspects of life, including the workplace. Companies must offer equal employment opportunities to PWD and ensure their access to \textit{reasonable accommodations}, including assistive technology, in the workplace. Despite the ADA, PWD continue to face challenges and discrimination in the workplace. Invisible work is required to work around inaccessible work practices \cite{branham_invisible_2015} and individuals may feel the need to disclose their disability to avoid being perceived as incompetent or lazy \cite{branham_invisible_2015}. Due to concerns about negative career implications~\cite{shinohara_2011}, some PWD may not disclose their disability to their employer. However, when disability is not disclosed, employers are not required to provide accommodations to PWD, leading to further invisible work of self-accommodation \cite{branham_invisible_2015}. Beyond the USA, other countries such as the UK \cite{UK_Rights_2016} and India \cite{India_Rights_2016} have also enacted similar laws to protect the rights of PWD.
 
\paragraph{\textbf{Work Experiences of BLV Professionals}}
Software professionals spend much time collaborating in meetings~\cite{stray_planned_2018}, and while hybrid meetings can be beneficial due to the flexibility for attendees to join from home with their customized work setup~\cite{alharbi_hybrid_2023}, hybrid meetings too are problematic for BLV professionals. Following presentations can be difficult~\cite{tang_understanding_2021}, as screen readers are unable to interpret the content on a shared screen. Hosting meetings for BLV facilitators is challenging for many reasons, including the inability to simultaneously follow the textual chat (with a screen reader) and the audio of a meeting \cite{akter_facilitator_2023}. These accessibility issues lead to BLV individuals feeling unable to fully contribute in meetings \cite{das_it_2019} and feeling that they are perceived as being incompetent as a leader or facilitator, leading to concerns on how these perceptions could harm their careers \cite{akter_facilitator_2023}.

Within the BLVSP population, the majority of accessibility studies involving BLVSPs have focused on coding-related tasks \cite{mountapmbeme_litreview_2022}. The visual nature of IDEs can cause difficulties in common coding activities such as code navigation \cite{albusays_interviews_2017}, discoverability, glanceability, and alertability \cite{potluri_codetalk_2018}. For example, screen readers verbalize an indentation as a sequence of spaces. This makes it hard for BLVSPs to navigate and read the code in languages such as Python. To get past these issues, extra work is required, e.g., by utilizing workarounds such as the use of print statements for debugging rather than using the IDE's debugging tool~\cite{albusays_interviews_2017} or writing a custom script for the screen reader~\cite{albusays_interviews_2017}. Fully participating in pair programming~\cite{hannay_effectiveness_2009} with sighted colleagues is also difficult for BLVSPs, since sighted \say{drivers} do not have AT installed on their machines, so BLVSPs cannot follow along as the \say{navigator}. To overcome some of these issues, solutions involving custom IDE plugins that work with screen readers have been proposed~\cite{armaly_audiohighlight_2018, potluri_codetalk_2018, potluri_codewalk_2022}.

In addition to IDEs, many other tools and resources are used to complete tasks, although they vary in accessibility. Coding tasks often require developers to seek information from sources such as blogs, tutorials, and forums. These websites are often inaccessible, requiring custom setups and additional work to access~\cite{storer_IDE_2021}. Building and deploying software requires developers to interact with Command Line Interfaces (CLIs), which are inaccessible with screen readers due to their unstructured text. This results in workarounds such as manually copying the output into a text editor where it can be more easily navigated by a screen reader~\cite{sampath_CLI_2021}.

Collaborative activities, such as software design and UI design \cite{pandey_understanding_2021} and project management \cite{huff_workexp_2020}, are heavily reliant on the display of and manipulation of visual information. Screen readers often cannot parse or interact with this visual information, which makes it difficult for BLVSPs to fully participate in the activity. Collaborating via shared documents is also challenging as screen readers struggle to understand some of the visual markers commonly used to indicate authorship or versioning, requiring workarounds such as using a screen reader on the different versions of the document and mentally making note of the changes~\cite{das_it_2019}.


Collectively, the existing research on workplace  experiences of BLVs in general and BLVSPs specifically illustrates the accessibility barriers faced that impact one's ability to complete work tasks. Many of these difficulties stem from inaccessible software development tools requiring additional work in the form of workarounds and customization. However, these studies do not explore how this additional work and inability to fully contribute in collaborative work impacts the career mobility of BLVSPs. This paper addresses this gap. We thereby address the call made by Albusays et al. on the diversity crisis of the software industry~\cite{albusays_diversity_2021}, specifically by understanding the perspectives of this underrepresented group toward career mobility.

\paragraph{\textbf{Career Mobility in Software Engineering}} 
The two most common career paths in SE are the technical path and the managerial path~\cite{tremblay_determinants_2002}. Individual factors (e.g., the desire for promotion)~\cite{tremblay_determinants_2002} and external factors (e.g., organizational, industrial, societal)~\cite{alavi_qualitative_2012} can affect the career path chosen by a software engineer. Additional factors may impact the career mobility for underrepresented groups working in the technology industry, yet there is little research on career mobility and underrepresented groups in SE, with the exception of women in SE. Ahuja
~\cite{ahuja_women_2002} identified various structural determinants (i.e., lack of access to informal networks, lack of mentors, institutional structures) that impact career progression of women. Armstrong et al.~\cite{armstrong_advancement_2018} extended Ahuja et al.'s model with social determinants (i.e., social expectations, work-family conflict) that may also affect the careers of women in the U.S. technology industry. Our paper addresses the gap that the career mobility of BLVSPs have not been considered in SE diversity research.

\section{Study Design} \label{study_design}

\begin{center}
\begin{table*}[ht]
\tiny
\resizebox{0.95\linewidth}{!}{
\begin{NiceTabular}
{|c|c|c|c|c|c|c|}
\toprule
ID & Job Position & Experience (years) & Organization Type & Self-Disclosed Vision Status & Age & Gender\\ \midrule
P1  & Software Engineer & 1 to 5 & Finance & light perception & 26 to 30 & M\\ \hline
P2  & Software Engineer & 6 to 10 & Media & totally blind & 31 to 40 & M\\ \hline
P3  & DevOps Engineer   & 1 to 5 & IT & legally blind w/ tunnel vision & 31 to 40 & M\\ \hline
P4  & Consultant        & 6 to 10 & Consulting & 2\% vision & 31 to 40 & M\\ \hline
P5  & Software Engineer & 1 to 5 & IT & low vision & 26 to 30 & W\\ \hline
P6  & Business Owner    & 1 to 5 & Non-profit & legally blind & 20 to 25 & W\\ \hline
P7  & Accessibility Specialist & 1 to 5 & Finance & completely blind & 20 to 25 & M\\ \hline
P8  & Software Engineer & 6 to 10 & IT & totally blind & 26 to 30 & M\\ \hline
P9  & Technical Executive * & 16+ & Finance & blind/low vision & 51+ & M\\ \hline
P10 & Accessibility Specialist & 6 to 10 & Finance & some light perception & 51+ & M\\ \hline
P11 & Software Engineer & 1 to 5 & IT & 20/150 corrected, blurry & 26 to 30 & M\\ \hline
P12 & Product Manager * & 11 to 15 & IT & completely blind & 31 to 40 & M\\ \hline
P13 & Software Engineer & 16+ & IT & totally blind & 51+ & M\\ \hline
P14 & Solutions Analyst & 1 to 5 & Higher Education & limited field vision & N/A & W\\ \hline
P15 & Software Engineer & 6 to 10 & Non-profit & totally blind & 31 to 40 & M\\ \hline
P16 & Accessibility Specialist & 1 to 5 & IT & totally blind & 31 to 40 & M\\ \hline
P17 & Accessibility Specialist * & 16+ & Finance & completely blind & 51+ & M\\ \hline
P18 & Accessibility Specialist & 6 to 10 & IT & totally blind & 31 to 40 & M\\ \hline
P19 & Software Engineer & 6 to 10 & Outsourcing & totally blind & 31 to 40 & M\\ \hline
P20 & Software Engineer & 1 to 5 & IT & totally blind & 26 to 30 & M\\ \hline
P21 & Software Engineer & 1 to 5 & IT & totally blind & 20 to 25 & M\\ \hline
P22 & Accessibility Specialist * & 16+ & IT & visually impaired & 51+ & M\\ \hline
P23 & Accessibility Specialist * & 16+ & IT & completely blind & 51+ & M\\ \hline
P24 & Software Architect & 11 to 15 & IT & low vision & 31 to 40 & M\\ \hline
P25 & Accessibility Specialist & 6 to 10 & IT & nearly blind w/ shape/color perception & 31 to 40 & M\\ \hline
P26 & Accessibility Specialist & 1 to 5 & Healthcare & totally blind w/ light perception & 31 to 40 & W\\ 
\bottomrule
\end{NiceTabular}
}
\caption{Detailed information of participants. For participant anonymity, all participant names were replaced with IDs, age is reported in ranges, and job titles do not include specific position information. The asterisk * signifies positions of management.}
\label{tab:demographic}
\end{table*}
\end{center}

We utilized semi-structured interviews as our research method, since its open-ended nature provides flexibility to explore topics of interest~\cite{seaman_qualitative_2021}. Interviews were conducted between June 2023 and August 2023 by the first author, and lasted between 53 and 108 minutes with an average of 76 minutes. No pilot interviews were conducted. Some questions were predefined and others were improvised in the flow of the conversation. Topics on the accessibility of software development meetings and their strategies to handle access challenges, career goals, and perspectives on career mobility were covered in the interviews. Predefined questions included questions such as \say{Do you ever spend time outside of meetings, handling things related to your meetings?}, \say{Where do you want to ideally see yourself in 10-15 years in your career?}, and \say{Do you think the accessibility issues we discussed today have any impact on career advancement?} Through personal contacts, snowball sampling, and online discussion groups (e.g., program-L), we recruited and interviewed 26 software professionals (\autoref{tab:demographic}) who self-identified as either being blind (n=14) or having low-vision (n=12). Participants held a range of software development roles (e.g., programmers, DevOps, testers, product managers), had at least one year experience in their respective position at the time of the interview, and were located in the U.S., Europe, or India. 

The study was approved by the researchers' Institutional Review Board. Participants were emailed a study information sheet prior to the interview. Verbal informed consent was collected at the start of the interview as well as permission to record the interview. Interviews were audio recorded and transcribed, with any identifying information anonymized, or broadly categorized as appropriate in the transcriptions prior to analysis~\cite{pandey_understanding_2021}. Participants were compensated at a rate of \$40 per hour.

We conducted inductive analysis~\cite{charmaz_grounded_2006} on the interview transcripts, including open coding~\cite{charmaz_qualitative_2012}, constant comparison~\cite{glaser_comparative_1965}, and memo writing, to develop themes that captured recurring patterns. Two authors individually analyzed the first five transcripts and met several times throughout the process to identify initial themes, with the remaining 21 transcripts analyzed by one author each. As new themes emerged, the researchers re-analyzed prior transcripts for those themes. Codes and sub-codes generated through our analysis mapped directly onto headings and subheadings in the Findings Section \ref{sec_findings}. 

\section{Findings} \label{sec_findings}
To answer our research question, \say{\textit{What factors affect career mobility of BLVSPs?}}, we discuss four factors that emerged from our data:
\begin{enumerate}
    \item \textbf{Technical Challenges:} Software development tools have usage challenges for BLVSPs, which inhibits full participation of BLVSPs in collaborative work such as meetings.
    \item \textbf{Perceptions of Colleagues on BLVSPs:} The behaviors and attitudes of colleagues can undermine career mobility.
    \item \textbf{Perceptions of BLVSPs about Management Roles:} Individuals have differing opinions on the accessibility of managerial work, impacting their career path choices.
    \item \textbf{Accessibility Investment:} Individuals invest time and energy to make their role accessible. This can lead to a reluctance to move within or across organizations.
\end{enumerate}

\subsection{Technical Challenges} \label{findings_4.1}
Software professionals rely on a variety of tools to complete their tasks \cite{jackson_collaboration_2022}. While progress has been made to improve the accessibility of these tools--e.g., \myparticipantquote{Figma is at least making headway into building things to be accessible}{P18}--our participants highlighted ways in which accessibility and usability issues arising from software, hardware, collaborative practices, and technology procurement have the potential to negatively impact the careers of BLVSPs.

\subsubsection{Inaccessible Software} 
Aligned with previous research~\cite{pandey_understanding_2021}, our participants discussed difficulties using many of the popular developer tools when collaborating with teammates, such as project management software (e.g., Jira, Trello), diagramming and whiteboarding software (e.g., Miro, Bluescape, LucidChart), and design tools (e.g., Envision). Our participants noted that, while tools are technically accessible (i.e., the tools implement recommendations from the Web Accessibility Initiative \cite{WAI}), they have usability issues for BLVSPs which can slow them down and hinder their participation in collaborative work. One common issue was due to the technical implementation of the screens causing navigation problems with screen readers: \myparticipantquote{When everything is just clickable text, I have nothing to navigate by.}{P1}. 
The usability of a tool for BLVSPs was not binary; it was dependent on the task the user was trying to perform. For example, one participant noted that, Jira was usable for their work: (\myparticipantquote{I change the status of tasks from completed to QA or assign to someone else... So they are accessible.}{P19}), whereas another participant noted it was difficult to use the Jira backlog screen in planning meetings, since the screen is hard to navigate: \myparticipantquote{It gets frustrating not being able to quickly jump down through the pseudo-table to a specific issue.}{P18}

In-house tools were also identified as having accessibility problems. One participant noted that they could not use an internally developed dashboard that collated information from multiple systems, such as build systems, requiring them to ask a colleague for assistance, \myparticipantquote{There's a lot of images on it. There's not a lot of other information. There's not a lot of impetus for making an accessible one. And I typically will just get somebody else on the team to read that information.}{P21}

\subsubsection{Inaccessible Hardware} 
Although hybrid work, where some attendees are in the room and others join remotely, is a reality for many \cite{smite_future_2023}, we found that the meeting room infrastructure supporting such hybrid meetings can be inaccessible for BLVSPs. As P12 noted, \myparticipantquote{All of our meeting rooms that are set up for virtual meetings have a touch screen that you have to interact with in order to do it. And it's a locked-down system that you can only interact with the touch screen. There's no keyboard, there's no chance to load on JAWS, there's nothing.}{P12}. This meeting room infrastructure thus erects a barrier for BLVSPs to act as effective facilitators or leaders of a meeting, adding to previously-documented digital barriers \cite{akter_facilitator_2023}. 

\subsubsection{Inaccessible Collaboration Practices} \label{finding_inaccessible_collab_practices}
The accessibility issues with workplace software and hardware were amplified in synchronous meetings where much collaborative work within software teams occurs \cite{stray_planned_2018}. Participants experienced situations in which they had to ask others for help when contributing, had to devise ways to contribute in alternate ways to the rest of the team, or, in some cases, could not contribute at all. For example, when discussing how to contribute to a digital whiteboard tool used in sessions such as retrospectives or planning meetings, P18 asked a manager to \say{\textit{just email them and just sa[y], `Hey this is what I ran into,' and they put it up on the board for me.}} On the other hand, P2 did not use the tool: \say{\textit{I tend to just verbalize my thoughts,}} while P1 \say{\textit{just kind of had to listen to the meeting.}}

Additionally, as noted in \cite{tang_understanding_2021}, screen readers do not work on shared screens (e.g., when presenting slides or pair programming), requiring BLVSPs to depend on their colleagues for assistance. When colleagues do not assist, difficulties arise, leading to inability to participate. Two participants noted common anti-patterns from colleagues were lack of verbalization (e.g.,\myparticipantquote{...[colleagues] would not be verbalizing what was on display,}{P26}) and moving around the screen too quickly (e.g., \myparticipantquote{People tend to scroll around a lot. And it becomes very difficult to follow}{P3}).

Though meeting accessibility sometimes relied on colleagues, some participants noted they felt uneasy asking for support. They felt this would lead to misperceptions about BLVSP not being as competent as a sighted person, with a negative impact on their career: \myparticipantquote{...and these misconceptions can be detrimental to your prospects in the company.}{P16}

Overall, the difficulties in contributing to meetings resulted in our participants feeling \myparticipantquote{frustrated}{P18}, \myparticipantquote{uncomfortable}{P20}, with a sense of \myparticipantquote{separation and exclusion}{P26}, sometimes even leading to them to choose to \myparticipantquote{check out and leave}{P25}. 

These negative feelings are concerning, as our participants were cognizant that fully contributing to meetings is helpful for demonstrating their professional abilities to peers and senior colleagues and thus \myparticipantquote{progress[ing] in your job}{P26}. Being unable to fully participate raised concerns amongst our participants about the potential impact this could have on their careers. As noted by P18:
\begin{quote}
    \textit{``If you're barred from participating in the meeting and people don't know who you are, or you can't assert influence, you just become a quiet person in the corner... I can see that being a direct barrier to advancement.''}
\end{quote}

\subsubsection{Ableist Technology Procurement Policies}

Multiple participants (i.e., P2, P12, P18, P20, P25, P26) shared that more accessible technology for BLVSPs could be provided if the procurement teams in organizations were aware of accessibility needs and incorporated these needs into the organization's procurement process: \myparticipantquote{I think there needs to be education on the procurement side.}{P12} Unfortunately, even when accessibility procurement protocols are in place, adherence is inconsistent and not applied to engineering teams: \myparticipantquote{The policy is just completely ignored... [if they] see a tool that [they] think is going to be useful.}{P2} 

\subsection{Software Professionals' Perceptions of BLVSPs}
Our participants overwhelmingly expressed that ableist assumptions and misunderstandings held by other software professionals impact career mobility of BLVSPs. Their experiences of inaccessibility in the workplace led to colleagues undermining career mobility and a systemic lack of representation of BLVSPs in managerial positions. 

\subsubsection{Misperceptions of BLVSPs}
Our participants dealt with persistent ableist assumptions of sighted colleagues that ``propagate" through their organizations and the software industry. P4 shared, \textit{``There's a certain amount of discrimination against blind people in the sense that people really do have very low expectations of blind people.
''} P16 mentioned the widespread misconception among sighted individuals that \textit{``a blind person cannot be a software developer or software engineer''} (P16), which makes asking for accommodations more daunting, due to the fear of giving credence to this false perception and being subject to inequitable treatment from colleagues. 

When \textit{peers} were aware of one's vision disability, there was a risk they might \textit{``try and be kind [to me] and... take something [tasks] that they think would be difficult for me [away].''} (P12). This reduced opportunities to demonstrate one's skill and contribute like any other team member. Misperceptions made it difficult for BLVSPs to openly address real--as opposed to assumed--accessibility issues with teammates, affecting one's ability to ``go up the hierarchy'' in the company:
\textit{``If it's a bit challenging for you to let your team and your organization know about the challenges that you are facing, then it might get a bit tricky getting those promotions or going up the hierarchy.''} (P20)

When \textit{managers} were aware of one's vision disability, they had outsized power to either ameliorate or exacerbate ableism in the workplace.  Within the team, leadership could intervene in ableist behaviors of peers. For example, P9 shared, \textit{``If I ask for something, and it doesn't happen, or people just refuse, right? Then a little bit of push from my boss, or some other person in the department to sort of say, `We expect you to include [P9] in your meetings.'~''} Unfortunately, some managers diminished the BLVSP's role by preemptively assigning tasks to sighted peers, depriving BVLSPs' agency: \textit{``If [my previous manager] thought [a task] would be difficult for me, he would just hand it off to somebody else. Without consulting me, and usually right in front of me... it made me angry, right? That's my decision to say if I need help or not.''} (P12). Thus, this is \myparticipantquote{the perfect way to propagate ableist ideas... Able-bodied people deciding what a blind person can and can't do.}{P18}.

In the \textit{software industry} as a whole, misperceptions held by software professionals were considered a hindrance to career progression. First, when software professionals at other companies assume PWD are unable to participate in the industry, the result is inaccessible software development tools:
\begin{quote}
    \textit{``These software companies that are building these tools, they make them inaccessible. They always think about the business metrics, there's always that one project manager who is like `How many disabled people are actually using our software?' And it's zero, because we can't use it, then they think `Ok, well no one is using our software so we don't have to worry about it.' ''} (P18)
\end{quote}
P18 referred to this as a a \textit{``catch-22''}, in which PWD do not use their product due to inaccessibility, and companies neglect accessibility because PWD do not use their product.

Second, when software professionals assume BLVSPs are unable to be effective leaders, they are the last considered for promotions. While P12 wants to rise in the corporate ladder, he commented, \textit{``there's going to be a point where that's going to be a fight that I either have to take on, or I have to pivot.''} He projected that, although the number of women CEOs and vice presidents in the tech industry is increasing, there would be more barriers as BLVSPs: \textit{``I think that you're going to hit a lot more barriers, especially as you get into more director-level, vice president roles that people have a `type' that they kind of associate [with those roles].''}  

\subsubsection{Addressing Misperceptions through Self-Advocacy}
Our participants indicated that career mobility is affected by self-advocacy. P12 commented, \textit{``it is very heavily based on the individual, because I think there's a level of confidence and assertiveness that you need to have to overcome any other misconceptions that other people might have about how well you can do your job.''} While speaking up for oneself \textit{``can be intimidating when you're younger or just starting out to speak up''} (P25), P9 insisted that BLVSPs should learn how to advocate for themselves, even if they do not yet have the authority of a senior employee:
\begin{quote}
    \textit{``if you're newer, two things happen: one is that you have to learn how to ask for things that you don't feel comfortable asking for, because what if they say no? … So, you have to learn how to ask for those things. And the other thing is that you have to be prepared to stand up to people who are like, `I don't have time for that'... So, it takes effort, and it takes practice, and it takes some fortitude.''}
\end{quote}
While BLVSPs would not have had to fight against propagating ableism had they been perceived just as capable as sighted professionals, they had to prove their competence to their sighted colleagues. P9 shared: \textit{``I have to show up and prove to them that I can do the work.''} 

According to our participants who were in managerial positions, being in management granted them the power to advocate for accessibility in many ways, including engaging with senior executives. P22, with his seniority in the organization, is able to schedule meetings with executive staff (e.g., VP, senior VP) to impress on them the importance of accessibility in their organization:
\begin{quote}
    \textit{``I do make a point to use my role to talk to senior leaders of the company... I feel it’s my responsibility to make them understand the role of accessibility and tell the importance of it and I’m able to get those meetings, which maybe others would not be, more in a lower level grade, you may be more intimidated just maybe not thinking it’s your part of your job.''}
\end{quote}
Further, P22 tactfully appeals to executives' sense of priority regarding accessible technological innovations:
\begin{quote}
    \textit{``I'm building advocacy rather than trying to percolate up a big negative problem that has really negative energy associated with it. It's visionary basically, I'm talking about a future and what I'm typically saying is a future of technology that is more inclusive and more accessible to everyone. What is [the company]'s role in building that future and basically being very, very clear with the idea that the future is–that the world is changing, and we have to be a part of creating that new world or we're going to watch someone else do it.''}
\end{quote}

Being in management positions and having seniority and authority enabled our participants to speak up about accessibility. Aligning with P13's comment that \textit{``the higher up you are in the organization, the more ... you feel safe in saying, `Hey, I can't see this. Can you clarify?' ''}, P9 confirmed that his seniority and authority gives him a bigger voice: \textit{``I'm at this point, a senior contributor, and so have a big voice. And so, I get listened to.''}

\subsubsection{Underrepresentation of BLVSPs in Management and Perpetuating Misperceptions}

While being in management as a BLVSP means that they have more power to speak up about inaccessibility, participants reported a lack of BLVSPs in management and the resulting lack of mentorship: \textit{``At the tech giants, I can't think of any blind person that would be in charge of an entire branch, or department... I don't know of any mentors, or anybody that's been in that situation to kind of learn from, if that makes sense.''} (P12)

Further, participants felt that even if they progressed into management, colleagues could exploit the blindness against them, because of the lack of BLVSPs in management: 
\begin{quote}
    \textit{``management can be kind of cut throat... There's not that many blind exec[utive]s... In management often you'll have enemies. And I've been a mid-level admin and had enemies... If you're blind that's a big weakness... People can use that to undermine you.''} (P4)
\end{quote}

In addition to the lack of BLVSPs in upper level positions overall, when BLVSPs reach higher management positions, they are more likely to inhabit a role related to accessibility, that is, BLVSPs who are in higher management typically work in the department of accessibility: \textit{``Usually, it's somebody with a disability in charge of accessibility... I don't know how much of that is people just want to be able to say, `We have somebody with a disability in charge of our accessibility department.' ''} (P12) 
Even though BLVSPs desire to pursue careers in divisions other than accessibility, some may be pigeonholed into the role of accessibility due to companies' virtue signaling.

Some participants were not necessarily interested in pursing careers in accessibility, although they ended up working on accessibility to gain access in the workplace. For example, P4 mentioned, \textit{``If I wasn't blind, I would not probably be into accessibility. So I wind up doing it because I have to know about it because I have to do stuff for myself... But I want to be involved in other conversations and other projects and stuff that aren't related to those issues.''} Similarly, P12 shared: \textit{``I'm not particularly interested in accessibility as a career...I've done some work in it, but it's not going to be my first choice of a career. And so, I'm wondering how much that will limit me, eventually.''}

\subsection{Perceptions of BLVSPs about Managerial Progression} \label{findings_4.3}
Our participants had different motivations for progressing into the managerial track: some participants felt that management would be more challenging, whereas others felt the workload would be feasible. For P9, a technical executive, management brought \textit{different} challenges around effective communication with others, not \textit{more} challenges: \textit{``So, as the challenges change, so will your accessibility needs. But it's not like being a manager is a harder push for a blind person than being an individual contributor.''} However, other participants reported varying desires to stay in the technical track or advance into management, as detailed in Section \ref{findings_4.3.1} below. %

\subsubsection{Career Aspirations of BLVSPs} \label{findings_4.3.1}
Our participants had career aspirations that ranged from seniority in technical roles, gaining experience in mentoring, and management. While some participants \textit{``definitely think about management a lot and think I may go that path just because I enjoy mentoring and helping people''} (P24), others wanted to remain in technical roles. P2 reported wanting to gain experience and seniority in software development before considering a management position: 
\begin{quote}
    \textit{``I'd like to be senior… And then, I mean, it really depends, right? Because I don't want to accidentally go into management. If I ever feel like I'd be a good engineering manager, I'll become an engineering manager. I've managed a couple of interns in the past. I think if I was in the right team, like probably not this one, but after senior, I might consider product.''}
\end{quote}
In addition, P9 wanted to mentor and guide junior developers, while continuing to write code: \textit{``I want to do one or two more large development projects, because I think keeping your hand in on the actual development side of things makes you better at the product management and mentoring side of things.''} He was not at all interested in being a \textit{``people-manager''}.

Many BLVSPs viewed the extensive number of managerial meetings negatively and were concerned about the additional stress involved with management. Such concerns put them off the managerial track. 
For example, P1 and P5 recalled their colleagues in managerial positions being in meetings \textit{``all the time’’} (P5). P1 watched his developer colleague become a team lead, which appeared \textit{``stressful''}:
\begin{quote}
    \textit{``I've watched my [colleague], she was a senior developer, and she stepped in to be our team lead for a time... It looked very stressful. It did not look like I wanted to do that, ever... Nothing about it sounds appealing.''}
\end{quote}
 
\subsubsection{Management Perceived as Challenging}
Some participants expected management positions to be harder, even \textit{``overwhelming''} due to their perceptions of accessibility of the role. P3, who lost his vision a few years prior, expressed concerns about his vision status impacting performance as a manager:
\begin{quote}
    \textit{``I thought about management... And I definitely have concerns about whether or not I could do that effectively. Just with my vision. I think my life will be easier if I remain an individual contributor. Because if you think about it, from my perspective at least managers end up in a lot more meetings and... they present a lot more. And there is a time cost to me that's above and beyond what a sighted person has. That I have to prepare so much for these things. So it gives me pause. If not for the vision, I definitely would be up for it. But I definitely would have to really strongly consider it and see what it would entail.''}
\end{quote}
P3 pointed out that staying on the technical track and mentoring junior developers through pair programming could still be challenging but less so than management: \textit{``[mentoring] would still be somewhat of a challenge, but I don't think it would be so much that it would limit me. If I was a manager, I think it would be overwhelming.''}

The access issues encountered in meetings (Section \ref{finding_inaccessible_collab_practices}) influenced participants' beliefs in their abilities to take on management roles. 
Access challenges made participants believe that they would \textit{``have some trouble occupying more of like a full-time management or project management role''} (P11). P11 projected that \textit{``meeting-related pain points that I have would kind of become more frequent more during a meeting driven position,''} given that management positions are \textit{``meeting-driven''} (P11). Similarly, P26 commented: \textit{``meetings [in management] are more hands-on. If they're not accessible, then literally the work isn't going to get done.''}

P19 voiced concerns that being in management would result in more customer-facing meetings which could be challenging due to the unpredictable accessibility of materials clients bring in: 
\begin{quote}
    \textit{``They [management meetings] would certainly be harder to handle. Because many times the client comes with a design, it comes with a photo. And I would have to discuss it with some[one] – many of them would certainly find it weird that we get someone else to discuss. Or they need to explain supplemental things in the photo.''}
\end{quote}

\subsubsection{Management Perceived as Feasible}
In contrast to some participants' beliefs of management being more challenging, some participants perceived management as more viable, as they projected that meetings would be more conversational. Hence, meetings were described as \textit{``something that can be managed''} (P20), despite the increased quantity.
Similarly, P7 shared:
\begin{quote}
    \textit{``It's a lot more analytical work. And it's less design diagrams and mockups, and more budget spreadsheets and words. So, I think management itself could actually be a very accessible position for someone with blindness. And I will love to see more people get into management. I think it's somewhere that we can actually excel at with minimal modifications. At least modifications that aren't technical.''}
\end{quote}

In addition, participants projected that managerial positions would give them more control and a powerful voice regarding accessibility. P12 anticipated that being in senior management would allow him to determine accessibility dynamics: 
\begin{quote}
    \textit{``I think maybe more people would be conscious of not being so visual-heavy if I were the director, right? I think the smart people would be trying to get their point across and maybe not be as dependent on graphs if they know I can't see graphs. I don't want people to cater to me. I don't think they should have to, but I do think once you're the senior person in the room, people naturally cater to you and might even bring in accessibility.''}
\end{quote}


\subsubsection{Managers' Experiences of Management}

A total of 5 of our 26 participants (19.2\%) held managerial positions, and their experiences illuminated the realities of management roles. 
While participants in managerial positions confirmed some perceptions that managers have many meetings and participate in potentially inaccessible customer-facing meetings, BLVSPs in management reported that their meetings were more conversational and thus, more accessible.

Confirming that management is indeed a \textit{``meeting-driven role''} (P9), P12 shared, \textit{``I'm in meetings for probably five or six hours a day.''} Similarly, P9 described having many meetings while being a product manager: \textit{``When I was a product manager, my day was basically driven by my calendar. I had a few open spots during my day... I had lots of meetings.''} 
Corresponding to some perceptions of non manager-participants, being a manager meant being \textit{``much more customer-facing than when I was a developer''} (P12), where customers were more likely to bring inaccessible materials into the meeting. Sometimes managers had to sit in inaccessible meetings, such as those discussing User Interface (UI) functionality and aesthetics, because \textit{``unfortunately, there's nobody else that is in the same position, that works with the same teams that has that information.''} (P12) 

However, contrary to some non-managerial participants' perceptions, BLVSPs in management found most of their meeting-related experiences to be accessible and manageable. P22 described meetings as mostly being high-level and conversational rather than infrequent hands-on, \textit{``in-the-weeds''} working sessions, with visual materials that were \textit{``very difficult''} to access. 
Meetings were considered more accessible on the whole:
\begin{quote}
    \textit{``The nature of the time being spent with people... They're generally not meetings that we are sitting in front of spreadsheets and updating things and or using online tools that are like Jira or something like that. %
    It's not hands on, or other hands on updates, or very visual... 
    So they are conversations. It's the art of the conversation that is the meeting. It's the `What is my objective? What outcome am I looking for? How am I gonna get there?' But it's a conversation... Inherently conversations are pretty accessible to people with visual disabilities.''} (P22)
\end{quote}

While some participants might consider progressing into management positions where they can remain hands-on with the technical skills they have acquired in their roles, an increased number of meetings and their preferences against meetings could deter BLVSPs from aspiring to enter management positions.%

\subsection{Accessibility Investment and Career Mobility}
As BLVSPs gained work experience, they were more likely to hesitate to move positions within or across organizations, due to the time and energy invested in making their existing role accessible.

\subsubsection{Labor of Cultivating an Accessible Workplace}
According to our participants, accessibility accumulated as they became established in their roles. For example, P12 shared, \textit{``The longer you've been in a position, the more accessible it potentially becomes because you have the tools and you have the resources that you've already built up.''}  Additionally, P14 benefited in accessibility from memorization over time at work: \textit{``Because I was with that institution for so long I was able to memorize a lot of the things... that made me very effective.''} P18 shared that over time, his team started \textit{``baking in''} accessibility into meetings and seamlessly normalized accessible meeting practices: \textit{``As people know me and they know the other disabled folks that are there they'll start baking it in. They don't even have to think about it anymore, they just do it naturally at the start, being accessible and inclusive.''}

Multiple participants described having to make a large time investment up front to make their jobs accessible. P4 shared:  
\begin{quote}
    \textit{``Everything I do requires extra time... the extra time could take different forms. A lot of times the extra time is baked into me having sat for hours, weeks, months, years, really setting up tools and environments that I can use to make the footing a little more equal... No amount of preparation ad-hoc would allow me to read a Google Document that someone was sharing on their screen over Zoom, except for the fact that I've sat around for nine hours at some point, seven years ago and built a tool to do it.''} 
\end{quote}


Gaining accessibility at work also entailed learning strategic approaches to addressing accessibility problems. P2 took the lead on accessible tasks that he could become skilled in, in order to pass off inaccessible tasks such as navigating Jira in meetings that could inhibit contribution in team meetings:
\begin{quote}
    \textit{``If there's a task I can't do, or like realistically I can't do that well, I try to find something else to lead on instead. [For] example: We've got analytics in our products. I would respectfully suggest that I probably know the most about all of the stats out of all of the engineers... because I wanted something to lead on and be good at, to help replace the stuff that I'm not so good at... I've worked in quite a few different teams, and obviously it'd be better if I didn't have to, but that's served me quite well.''}
\end{quote}

Moreover, P2 learned over time at his job that phrasing accessibility problems in a \textit{``non-accessibility way''} is oftentimes effective in convincing his team to adopt accessible tools as the norm in the workplace:
\begin{quote}
    \textit{``If there's a tool that you can't use, and you need to be able to use it, the way to convince the team to move off it isn't to plead the accessibility argument, often. Find something that's really bad about the tool, and then make something that does it better. And then everyone will just start using your version, anyway.''}
\end{quote}
However, doing so requires P2 to put in the labor of thinking about how accessible tools can be argued for without explicitly framing them as an accessibility problem.

\subsubsection{Reluctance to Change Positions to Protect Accessibility Investment}
Since workplace accessibility took a lot of time and labor to accrue, participants felt hesitant to find new positions within or outside of their organizations even if they wanted to seek other opportunities. BLVSPs gained \textit{``a bank of respect that I've built up already that I can call on''} (P12) as they became established and gained seniority in their role. P22 believed that \textit{``being comfortable in an organization when you have a disability is a reason to stay. It shouldn't be a valid reason but I think it is.''}  Similarly, P23 shared that he stayed with one company for a long time because he accumulated accessibility at his job, and transitioning to a new organization would take a long time:
\begin{quote}
\textit{``that[transitioning]'s going to take a long time for the employee to fit into the company because even like going through the interview process, you’re going to have to teach people onboarding, you're going to have to teach people the tools that you'll use, who knows if you can use them after you get hired.''}
\end{quote}
P2, along with other participants, reported that the upfront cost of inaccessibility in jobs deterred him from changing teams: \textit{``now that I'm quite familiar with all of the systems that we use, that isn't as much of a problem. If I move teams, it will be.''} 

Participants described having to assure their coworkers that their disabilities would not hinder their work, and if they switched jobs, they would have to convince a new set of people of their abilities. P12 explained the reason for not leaving the accumulated respect and accessibility behind and looking for a new position:
\begin{quote}
\textit{``I'd have to convince them, not only am I good at product strategy, but blindness would not be an issue. And honestly, it's something that would make me hesitate leaving my current position, because I know with my current position, I'm seen as not a blind person, but as a competent person in my role.''}
\end{quote}

These cumulative rewards and footholds in the larger organization would disappear if BLVSPs suddenly moved to a different position or company. P12 reported, when talking about leaving a previous position: \textit{``I think I would have changed jobs sooner if I had more confidence and belief that disability would not have an impact.''}, in addition to explaining that his disability \textit{``did play a factor into me hanging out on that team as long as I did.''}
\section{Discussion}
Scholars in software engineering have been investigating career advancement in the IT industry and the programming experiences of BLV developers. However, they have paid little attention to the 
aspects of work outside of pure development, such as the career mobility of BLVSPs. To the best of our knowledge, this is the first study to examine the social and technical factors that impact the career mobility of BLVSPs in the workplace. 


Similar to prior work, we found that BLVSPs faced access barriers to the independent use of popular software development tools, like Jira~\cite{filho_visual_2015, huff_workexp_2020, pandey_understanding_2021}. Additionally, we found numerous digital tools that are technically accessible but practically unusable, costing BLVSPs precious time–the ``invisible work" reported for BLV employees more generally~\cite{branham_invisible_2015}–and forcing them to seek sighted assistance~\cite{bigham_effects_2017, das_it_2019, akter_facilitator_2023}. We extend these results substantively by documenting how these \textbf{accessibility and usability issues, especially during meetings, hampered BLVSPs ability to contribute to their fullest potential and on equal footing with their sighted peers}. An otherwise productive BLVSP became a \textit{``quiet person in the corner''} (P18) or felt the need to skip meetings altogether (e.g., P24, P26), \textbf{which was perceived as being detrimental to getting noticed and moving up the corporate ladder}. Our participants' intuitions align with prior research that documents how interpersonal communication and networking, which are critical to career progression ~\cite{wolff_effects_2009, gibson_understanding_2014}, are mostly forged in meetings~\cite{samrose_meetingcoach_2021, bly_media_1993, fitter_closeness_2020}. We contend that such social and technical barriers have the potential to impact both technical and managerial career progression of BLVSPs, as meetings are such a common collaboration and coordination mechanism irrespective of job role~\cite{stray_planned_2018, mroz_science_18}.



Interestingly, the \textbf{(mis)perceptions of sighted colleagues} about the accessibility of software development tools and tasks were as critical to career growth potential, if not more so than actual accessibility. In other words, ableist attitudes and unconscious biases of both peers and managers meant that BVLSPs were often cut out of the loop, removing opportunities to contribute and demonstrate their skills. Our findings contribute a disability-focused perspective to the current body of literature about biases against minoritized groups, such as women~\cite{michie_barriers_2006, yi_implicit_2019}, being less capable than men in the tech industry.


Even \textbf{BLVSPs themselves had (mis)perceptions} about what positions they could access, which could have detrimental career implications. Some BLVSPs assumed meeting-heavy managerial positions would \textit{not} be accessible, while others believed they would be \textit{more} accessible. The reality, according to our BLVSP managers, was that \textit{both} were true, depending on the job duty. This mismatch between perception and reality introduces two problems. First, some BLVSPs may not explore the option of managerial progression, choosing to remain in the technical track despite their interests. Second, BLVSPs may be entering into management positions without being cognizant of the challenges they may face. The disconnect may also signal a lack of direct communication and mentorship between junior and senior professionals with disabilities, which has been identified as crucial for disabled professionals~\cite{daughtry_mentoring_2009}. Though ours is the first study of career mobility to center BLVSPs, our findings parallel those about career barriers faced by  non-disabled women in technology, such as the scarcity of mentors in higher level positions~\cite{ahuja_women_2002}.

Finally, we found that BLVSPs spent an extraordinary amount of time and energy \textbf{cultivating an accessible workspace}. They negotiated accessible practices with colleagues, advocated for more accessible technology procurement, and spent \textit{``hours, weeks, months, years setting up tools and environments''} (P4) to make them effective workers. This ``accessibility investment,'' as we call it, is often so substantial and the risk of losing that investment so great that many BLVSPs felt locked in their current positions and perceived switching positions within or across companies as unmanageable. This is particularly concerning, as non-disabled IT professionals are known to switch jobs between different companies as a way to make a higher salary~\cite{naresh_job_hopping}, with such ``job hopping'' contributing to significant wage growth~\cite{ariga_wage_2013}. Once again, we see that accessibility barriers in the software industry also pose barriers to career progression, potentially with severe financial repercussions.



In sum, we extend and contribute to our field's understanding of workplace accessibility and career progression by documenting how our BLVSP participants believed that: (1) technically accessible but not totally usable workplace technologies, (2) the misperceptions of peers and superiors on BLVSPs
, (3) the discrepancy between perceptions and realities of advancing into managerial positions
, and (4) the accessibility at work that our participants accrue over time, all impact their career mobility. 


\subsection{Implications for Tools used in Software Development}
We advocate for the integration of accessibility considerations in the creation of software development tools and use of collaborative tools (e.g., Slack, Zoom) in software development. BLVSPs work on developing software that are used by millions of people. Yet, during development, they are often stuck grappling with inaccessible in-house and commercial tools, or even inaccessible products that they have to work on. Designers and developers of such tools--including in-house tools--should thoroughly understand the basic accessibility needs of BLVSPs'–including how their assistive technologies interact with software tools–and their specific needs in the software development workplace, especially in collaborative work settings such as meetings. Active involvement of BLVSPs in the design and development phase of tools~\cite{schuler_participatory_1993} is crucial to improve the accessibility of software development tools. 

\subsection{Implications for Organizations}
BLVSPs must have access to mentors who understand disability, including mentors with vision disabilities, in senior-level positions in the tech sector. However, the burden of finding mentors currently falls on the already overburdened BLVSPs who are dealing with access problems. Thus, we recommend that organizations sponsor mentorship programs, as they have been proven efficacious in helping PWD more broadly~\cite{daughtry_mentoring_2009}. Organizations should also sponsor awareness-raising initiatives targeting non-disabled employees, especially in-house developers, including digital accessibility training. Stronger policies regarding accessible technology procurement and digital infrastructure investments are necessary, to level the playing field for BLVSPs in the software industry. 

\subsection{Implications for Research}

People with disabilities comprise the largest minority group in the USA~\cite{gov_diverse_2023}. Thousands of people with disabilities are employed by top tech companies~\cite{Microsoft_DEI_2023, Google_DEI_2023}, and professional computing organizations increasingly include disability in their public diversity statements~\cite{IEEE_diversity, ACM_diversity}. Yet, diversity-oriented research in SE predominantly investigates gender, age, and race/ethnicity, leaving disability out~\cite{silveira_systematic_2019, rodriguez_perceived_2021}. Disability should be considered a factor of diversity, and we urge SE researchers to conduct more comprehensive research that encompasses disability, to understand disability in SE and to foster inclusive SE environments and work experiences.

\section{Threats to Validity}
We discuss the credibility, representativeness, reliability, and transferability of our research as recommended by Creswell \cite{creswell_qualitative_2018}.

\paragraph{\textbf{Credibility}} The multi-disciplinary research team contained experts in SE and Human-Computer Interaction (HCI). Some members of the team also had extensive experience working in the technology industry. This multi-disciplinary team was thus able to provide varied perspectives on the themes emerging from the analysis of the data, helping to minimize potential biases during analysis. Peer debriefing  \cite{creswell_qualitative_2018} was used to discuss the emerging themes with the wider research team. Member checking  \cite{creswell_qualitative_2018} was also utilized; we presented our findings to a number of participants (n=6). Member checking determined that our findings resonated with their personal experiences of career progression.

\paragraph{\textbf{Representativeness}} Our study consisted of 26 participants with a risk they are not truly representative of all roles in the software industry. We have mitigated this risk by selecting a diverse set of participants from a variety of companies and industries, roles, years of experience, and levels of vision. There is also a risk that our sample size was not large enough to determine findings applicable to the broader population of BLVSPs, which is always a risk in qualitative work. There was enough commonality in the answers from our participants that we feel code saturation~\cite{hennink_saturation_2017} was reached and our results are credible. Additionally, this study initially focused on BLVSPs' experiences in meetings, but during the process it involved discussions around their career mobility within the context of meetings. It is therefore possible that we do not fully encapsulate the entirety of the career mobility of BLVSPs. Additional studies are warranted to holistically understand their professional career progression beyond the confines of software development meetings. 

\paragraph{\textbf{Reliability}} To mitigate risks related to inconsistencies in capturing and analyzing data that could result in unreliable findings, we ensured consistent data collection and data analysis processes as described in Section \ref{study_design}.

\paragraph{\textbf{Transferability}} Our participants worked in different industries and organizations, with participants based in the USA, India, and Europe. Our findings are likely transferable to BLVSPs in similar roles and organizations within those geographic regions. However, some of the findings relate to organizational and social factors which are affected by local cultural work practices. Moreover, countries differ in the legal protection designed to protect people with disabilities from discrimination. Further research is required to determine if our findings are applicable to BLVSPs in other regions.
\section{Conclusion and Future work}
In this paper, we captured the results of semi-structured interviews with 26 BLVSPs and identified four factors that have the potential to impact the career mobility of Blind and Low Vision Software Professionals: (1) technical challenges, (2) how colleagues perceive them, (3) their own perceptions of managerial progression, and (4) their investment in workplace accessibility. These findings point to critical implications for software development tools, organizations, and researchers. Our recommendations include further research about disability as a factor of diversity in software engineering and involving BLVSPs in the design and development of software development tools (including in-house tools) to ensure usability of such tools by BLVSPs. Future work should also uncover the challenges BLVSPs may face in pursuing different career options and lateral movements in the tech industry.

\section{Acknowledgments}

We would like to thank our participants for their involvement in our study and for providing insights and perspectives. This work was supported by the National Science Foundation (NSF) awards \#2211790, \#2210812, and \#2326489.

\bibliographystyle{ACM-Reference-Format}
\bibliography{sample-manuscript}

\end{document}